\documentclass[a4paper,11pt]{article}
\usepackage{pos}
\usepackage{amsmath}
\newcommand{\eprint}[1]{{\tt [arXiv:#1]}} 
\newcommand{\et}[2]{\eta_{#1}^{#2}}
\newcommand{\dd}{\textrm{d}}

\renewcommand{\logo}{\relax}

\title{Resurgent Lambert series from Feynman and beyond}

\author*[a]{David Broadhurst}
\author[b]{Daniele Dorigoni}

\affiliation[a]
{19 Milfoil Avenue, Milton Keynes MK14 7DY, UK}
 \affiliation[b]
{Centre for Particle Theory \& Department of Mathematical Sciences,\\
Durham University, Lower Mountjoy, Stockton Road, Durham DH1 3LE, UK}

\emailAdd{djbroadhurst@yahoo.co.uk}
\emailAdd{daniele.dorigoni@durham.ac.uk}

\abstract{Lambert series of the form $\sum_{n>0}a(n)q^n/(1-q^n)$
are ubiquitous in mathematical physics.
In particular, 2-loop sunrise and 3-loop banana Feynman diagrams yield
Lambert series with $a(n)$ of the form $\chi(n)/n^s$ where $\chi(n)$ is a Dirichlet character.
Resurgence concerns the singular limit as $|q|$  approaches 1. In the Feynman
cases we can control this limit, obtaining rapidly convergent expressions, since the
Lambert series are iterated integrals of holomorphic Eisenstein series twisted by
a character. We generalize this result, to include modular resurgent
structures found in topological-string observables.}

\FullConference{Loops and Legs in Quantum Field Theory (LL2026)\\
12-17, April, 2026\\
Bayreuth, Germany\\}

\begin{document}
\maketitle

\section{Introduction}

We begin with an identity for $\zeta(3)$ from Ramanujan's notebooks. It comes from
a more general  formula that lay there undiscovered and unproved for 50 years~\cite{Berndt,Grosswald}.
In Section~2, we show how this
relates the appearance of $\zeta(3)$ in the behaviour of a 3-loop Feynman integral~\cite{BKV}
at small and large momenta. In Section~3, we explain how this may be regarded as an example
of Cheshire cat resurgence~\cite{Dorigoni:21}, with exponentially suppressed terms
surviving when a perturbative series terminates. We also show that a 2-loop integral
benefits from this simplicity, thanks to the inclusion of a quadratic character in a Lambert series. 
In Section~4, we outline a more general result from~\cite{BD}, where we considered a pair of
characters and found the non-perturbative terms that resolve ambiguities of resummation.
Finally, in Section~5, we apply our analysis to computations of spectral traces in
topological string theory~\cite{Fantini:24,Fantini:24a,Fantini:25,Rella}. We find that
compactification on a Calabi-Yau manifold with $\mathbb{P}^{m,n}$ geometry
results in appearance of Lambert series twisted by characters with conductor $N=m+n+1$.
At $N=5$, we encounter both odd and even characters and hence an admixture
of a more difficult case, requiring Borel resummation,  with a simpler
case, exhibiting  Cheshire cat resurgence.

 \section{Ramanujan's transformation of the 3-loop banana}

Ramanujan's notebooks~\cite{Berndt}  give an efficient way of evaluating $\zeta(3)$ from the Lambert series in  
\begin{equation}
\zeta(3)=\frac{7\pi^3}{180}
-2\sum_{n>0}\frac{1}{n^3}\frac{q^n}{1-q^n},\quad q=\exp(-2\pi) < \frac{1}{353}\label{z3},
\end{equation}
which is the first of his identities 
\begin{equation}
\zeta(4k-1)=\frac{1}{\pi}\sum_{n=0}^{2k}
(-1)^{n+1}\zeta(2n)\,\zeta(4k-2n)
-2\sum_{n>0}\frac{n^{-4k+1}}{\exp(2\pi n)-1}\,.
\end{equation}
Similarly $\zeta(5)$ is rapidly evaluated by setting $k=1$ in the identity~\cite{BB}
\begin{align}
k\,\zeta(4k+1)&=\frac{1}{\pi}\sum_{n=0}^{2k+1}
(-1)^n\,(n-\tfrac12)\,\zeta(2n)\,\zeta(4k+2-2n)
\nonumber{}\\&{}
-2\sum_{n>0}\frac{n^{-4k-1}}{\exp(2\pi n)-1}
\left(k+\frac{\pi n}{1-\exp(-2\pi n)}\right)\,.
\end{align}
We can improve~\cite{Bfam}  upon~(\ref{z3}),  by evaluating Lambert series with
$q\le\exp(-12\pi/\sqrt{15})<\frac{1}{16879}$ in
\begin{gather}\zeta(3)=\frac{16351\pi^3}{330^2\sqrt{15}}
+\frac{2}{121}W_{12}
+\frac{1368}{605}W_{15}
-\frac{6}{605}W_{20}
-\frac{2924}{605}W_{30}
+\frac{342}{605}W_{60}\,,\\
W_k=\sum_{n>0}\frac{1}{n^3}\frac{1}{\exp(\pi kn/\sqrt{15})-1}.
\end{gather}

Ramanujan's value of $\zeta(3)$ came from a family of quasi-modular transformations of
\begin{equation}
T_s(\tau)=\frac12\zeta(s)+\sum_{n>0}\frac{1}{n^s}\frac{q^n}{1-q^n},\quad q=\exp(2\pi{\rm i}\tau)=\exp(-2\pi y)
\end{equation}
with $\Im\tau=\Re{y}>0$. This converges well at large $y$. To access $|q|\to1^-$, as $y\to0^+$,
we use a Fricke involution $y\to1/y$, when $s$ is odd. For example, Ramanujan obtained~\cite{Berndt,Chavan}
 \begin{align}
yT_3({\rm i}/y)&=\frac{\pi^3}{180}\left(y^2+5+\frac{1}{y^2}\right)-\frac{T_3({\rm i}y)}{y}\\
y^2T_5({\rm i}/y)&=\frac{\pi^5}{3780}\left(2y^3+7y-\frac{7}{y}-\frac{2}{y^3}\right)+\frac{T_5({\rm i}y)}{y^2}\\
y^3T_7({\rm i}/y)&=\frac{\pi^7}{56700}\left(3y^4+10y^2-7+\frac{10}{y^2}+\frac{3}{y^4}\right)-\frac{T_7({\rm i}y)}{y^3}
\end{align}
which originate from Eichler integration of holomorphic Eisenstein series
\begin{equation}
\sum_{n>0}\frac{n^s q^n}{1-q^n} =\left(q\frac{\rm d}{{\rm d}q}\right)^s\sum_{n>0}\frac{n^{-s}q^n}{1-q^n}
\end{equation}
with even modular weights $s+1\ge4$. Iterated integrals of Eisenstein series and other modular forms 
have been intensively studied in quantum field theory~\cite{Adams,Broedel} and in string theory~\cite{Dorigoni:22}.

The Schwinger parametrization  of the equal-mass 3-loop banana integral in two spacetime dimensions gives~\cite{BKV} 
\begin{equation}
J(t)=\int_0^\infty\int_0^\infty\int_0^\infty\left(\frac{1}{(1+a+b+c)(1+1/a+1/b+1/c)-t}\right)\frac{\dd a}{a}\frac{\dd b}{b}\frac{\dd c}{c}
\end{equation}
and for real $t<16$ it is given by the Bessel moment~\cite{BBBG}
\begin{equation}
J(t)=8\int_0^\infty I_0(\sqrt{t}x)K_0^4(x)x\,\dd x .
\end{equation}
Broadhurst and Laporta discovered the on-shell evaluation~\cite{BBBG}
\begin{equation}
J(1)=\frac{\pi^3}{2}\left(1-\frac{1}{\sqrt{5}}\right)
\left(1+2\sum_{n>0}\exp(-\sqrt{15}\pi n^2)\right)^4
=\frac{\Gamma\left(\frac1{15}\right)\Gamma\left(\frac2{15}\right)
\Gamma\left(\frac4{15}\right)\Gamma\left(\frac8{15}\right)}{30\sqrt5},
\end{equation}
which was later proved by work in~\cite{BKV}, completed in~\cite{Samart}.

To develop the Taylor series  $J(t)=\sum_{n\ge0}C_nt^n$, one may use the recurrence relation~\cite{BBBG,BorweinSalvy}
\begin{equation}
C_n=\left(1-\frac{1}{2n}\right)\left(5-\frac{5}{n}+\frac{2}{n^2}\right)C_{n-1}-4\left(1-\frac{1}{n}\right)^3C_{n-2}.
\end{equation}
for $n>1$, with initial conditions 
$C_0=7\zeta(3)$ and  $C_1=C_0-6$ that ensure that $J(4)$ is finite.
This recurrence relation determines the inhomogeneous third-order Picard-Fuchs equation
\begin{equation}
t^2(t-4)(t-16)J^{\prime\prime\prime}
+6t(t^2-15t+32)J^{\prime\prime}
+(7t^2-68t+64)J^\prime
+(t-4)J+24=0
\end{equation}
with singularities for $t\in\{0,4,16,\infty\}$.

A modular parametrization of the homogeneous equation was given by Helena Verrill~\cite{Verrill} in 1996, using
the Dedekind eta function 
\begin{equation}
\eta(\tau)=q^{1/24}\prod_{n=1}^\infty(1-q^n)=\sum_{n=-\infty}^\infty(-1)^n q^{(6n+1)^2/24}
=\frac{\eta(-1/\tau)}{\sqrt{-{\rm i}\tau}}
\end{equation}
where $q=\exp(2\pi{\rm i}\tau)$. Denoting $\eta_n:=\eta(n\tau)$, let
\begin{equation}
t=-64\left(\frac{\eta_2\eta_6}{\eta_1\eta_3}\right)^6,\quad
\psi=\frac{\et12\et32}{\et2{}\et6{}},\quad h=\frac{\et2{16}}{\et18}-9\frac{\et6{16}}{\et38}
\end{equation}
where $\psi^2$ is the square of a elliptic integral, obtained by Geoff Joyce~\cite{Joyce}  in 1973.
as a solution to the homogeneous equation, and $h$ is a weight-4 cuspform of level 6~\cite{Acres}.
Then we obtain the inhomogeneous equation
\begin{equation}\left(q\frac{{\rm d}}{{\rm d}q}\right)^3\frac{J(t)}{\psi^2}=
24h=24\sum_{n=1}^\infty\frac{n^3(q^n-8q^{3n}+q^{5n})}{1-q^{6n}}
\end{equation}
which is easily integrated to give Lambert series in
\begin{equation}
\frac{J(t)}{\psi^2}=H(\tau)=24T_3(\tau)-3T_3(2\tau)-8T_3(3\tau)+T_3(6\tau).
\end{equation}
The Lambert series converge well for $|t|\le8$ and at $t=0$ give the correct value of
\begin{equation} 
J(0)=\tfrac12(24-3-8+1)\zeta(3)=7\zeta(3).
\end{equation}

The Fricke involution
$\tau\to-1/(6\tau)$ gives $t\to64/t$ and Ramanujan delivers
\begin{equation}
48\tau^2H\left(\frac{-1}{6\tau}\right)=
4(2\pi{\rm i}\tau)^3+16(3T_3(\tau)-6T_3(2\tau)-T_3(3\tau)+2T_3(6\tau))
\end{equation}
for fast evaluation of $J(t)$ with $|t|\ge8$. Then, at large $|t|$, we obtain
\begin{equation}
-tJ(t)=4\log^3(-t)+16\zeta(3)+O(\log^3(-t)/t)
\end{equation}
whose asymptotic constant $16\zeta(3)$  is determined  by $J(0)=7\zeta(3)$.
Thus Ramanujan's notebooks, from more than 100 years ago,  provide us with a 
connection between expansions at large and small momenta. In Section~5 we use  
a Fricke involution to relate expansions at strong and weak couplings
for topological string observables. 

\section{Cheshire cat resurgence}
\begin{quotation}\noindent{\em
``Well! I’ve often seen a cat without a grin,'' thought Alice; ``but a
grin without a cat! It’s the most curious thing I ever saw in my life!''}
(Charles Lutwidge Dodgson, alias Lewis Carroll, in {\em Alice's Adventures in Wonderland}, 1865.)
\end{quotation} 

Ramanujan's very useful formula
\begin{equation}T_{2k+1}\left(\frac{-1}{\tau}\right)=
\frac{{\rm i}}{\pi}\sum_{j=0}^{k+1}\frac{\zeta(2j)\zeta(2k+2-2j)}{\tau^{2j-1}}+\frac{T_{2k+1}(\tau)}{\tau^{2k}}
\end{equation}
has a terminating Laurent series and an exponentially suppressed tail, as $\tau\to{\rm i}\infty$.
By contrast, Lambert series $\sum_{n>0}n^{-s}q^n/(1-q^n)$ with even $s$ give divergent series,
requiring directional Borel resummation whose ambiguities are eventually resolved 
by careful treatment  of the exponentially suppressed tail. 

In~\cite{Dorigoni:21} the quasi-modularity of Ramanujan's formula for odd $s$ was described
as an example of ``Cheshire cat resurgence'', since  exponentially suppressed terms remain after
the necessity of resummation of the perturbative terms has disappeared. 
Amusingly, we know of a Cheshire cat with even  $s$.  
The 2-loop sunrise diagram produces Lambert series with $s=2$.
So how can it avoid a divergent asymptotic series at large momenta?

For the sunrise integral $I(w^2)=4\int_0^\infty I_0(wx)K_0^3(x)x\,dx$, 
the modular parametrization~\cite{Acres,BV}
\begin{equation}
w=3\frac{\et22\et34}{\et14\et62}\,,\quad
f=\frac{\et16\et6{}}{\et23\et32}\,,\quad
g=\frac{\et39}{\et13}+\frac{\et69}{\et23}
\end{equation}
neatly reduces the problem to solving
\begin{equation}-\left(q\frac{{\rm d}}{{\rm d}q}\right)^2\frac{I(w^2)}{f}=6g
=6\sum_{n=1}^\infty\frac{n^2(q^n-q^{5n})}{1-q^{6n}}
\end{equation}
and the integration produces, inter alia, a Lambert series~\cite{BV} 
\begin{equation}
\mathcal{L}_2(\chi_{6,5};q)=\sum_{n=1}^\infty\frac{\chi_{6,5}(n)}{n^2}\frac{q^n}{1-q^n}
\end{equation}
with an odd character $\chi_{6,5}(n)=\pm1$ for $n=\pm1~\text{mod}~6$ and $\chi_{6,5}(n)=0$ otherwise.
Then quasi-modularity~\cite{Acres} allows one to expand about any of the singular points
$w\in\{0,1,3,\infty\}$ that locate cusps of modular forms on the congruence subgroup $\Gamma_0(6)$.

\subsection{Two types of divisor sums}

The Lambert series
\begin{align}
\mathcal{L}_s(\chi_r ; \tau)& =\sum_{n>0}  \frac{\chi_r(n)}{n^s} \frac{q^n}{1-q^n}\nonumber\\ 
&=\sum_{n>0}\frac{\sigma'_{s,\chi_r}(n)\,q^n}{n^s},\quad 
\sigma'_{s,\chi_r}(n)= \sum_{d|n} \chi_r(n/d)d^s,
\end{align}
has a Fourier expansion whose $n$-th coefficient involves sums of powers of divisors $d|n$ weighted
by the character $\chi_r(n/d)$. We shall also need the dual series  
\begin{equation}
\widetilde{\mathcal{L}}_s(\chi_r;\tau)= \sum_{n>0}\frac{\sigma_{s,\chi_r}(n)\,q^n}{n^s},\quad
\sigma_{s,\chi_r}(n)  = \sum_{d|n} \chi_r(d) d^s.
\end{equation}

If $\chi_D$ is the  primitive quadratic character with fundamental discriminant $D$,
then we have quasi-modularity, for odd $s$ when $D>1$,
and for even $s$, when $D<0$.  With conductor $N=|D|$, the modularity gap is given by evaluations of 
Bernoulli polynomials, with
\begin{gather}
\sqrt{D}\left(2\mathcal{L}_s(\chi_D;\tau)+\sum_{n>0}\frac{\chi_D(n)}{n^s}\right)
+2(-\tau)^{s-1}\widetilde{\mathcal{L}}_s\left(\chi_D;\frac{-1}{N\tau}\right)=\nonumber\\
\frac{(2\pi {\rm i})^s}{(s+1)!}\sum_{j=0}^{\lfloor s/2\rfloor}
\tau^{2j-1}{s+1\choose 2j}B_{2j}(0)\sum_{m=1}^{N-1}
\chi_{D}(m)B_{s+1-2j}(m/N).\end{gather}

\section{ The general case: resolving an ambiguity in resummation}

In~\cite{BD} we generalized by including pairs of characters and indices in
\begin{equation}
\Xi_{s_1,s_2}(\chi_{r_1},\chi_{r_2}; \tau)= \sum_{n_1=1}^\infty \sum_{n_2=1}^\infty 
\frac{\chi_{r_1}(n_1)}{n_1^{s_1}} \frac{\chi_{r_2}(n_2)}{n_2^{s_2}} q^{n_1 n_2}.
\end{equation}
With $\tau={\rm i}y$ and $y\to0^+$ we found that resummed perturbative terms 
require, in general, an exponentially suppressed tail that is a sum of transformed terms:
\begin{equation}
\sum_{n=0}^\infty \frac{ (s_1)_n (s_2)_n}{n!}\, \frac{(r_1 r_2 \tau)^{s_1+s_2+n-1}}{(-2\pi {\rm i})^n}
\,\Xi_{s_1+n,s_2+n}\left(\bar{\chi}_{r_2},\bar{\chi}_{r_1}; \frac{-1}{r_1 r_2\tau}\right)
\end{equation}
with only one term surviving in the single character cases
\begin{align}
\mathcal{L}_s(\chi_{r};\tau) &= \Xi_{s,0}(\chi_{r},\chi_{1,1};\tau)
= \Xi_{0,s}(\chi_{1,1},\chi_{r};\tau)\\
\widetilde{\mathcal{L}}_s(\chi_{r};\tau)& = \Xi_{0,s}(\chi_{r},\chi_{1,1};\tau) 
= \Xi_{s,0}(\chi_{1,1},\chi_{r};\tau)
\end{align}
relevant to work by Veronica Fantini and  Claudia Rella on
quantum modularity of resurgent topological strings~\cite{Fantini:24,Fantini:24a,Fantini:25,Rella}.

\section{Topological strings} 

These authors studied a
fermionic spectral trace, in topological string theory compactified on a Calabi-Yau manifold with $\mathbb{P}^2$ geometry,
obtaining $q$-Pochhammer symbols in
\begin{gather}\mbox{Tr}(\rho_{\mathbb{P}^2}) = \frac{1}{3 \sqrt{\tau}}
\exp\left(-\frac{\pi {\rm i} }{12} \tau +\frac{\pi {\rm i}}{ 36} \tau^{-1} +\frac{\pi {\rm i} }{4}\right)\,
\frac{(q^{2};q^3)_\infty^2}{(q;q^3)_\infty}\, \frac{(\omega_3;\tilde{q})_\infty}{(\omega_3^{-1};\tilde{q})_\infty^2},\\
(x q^\alpha;q)_\infty := \prod_{n=0}^\infty \left(1-xq^{\alpha+n}\right),\quad
q:=\exp(2\pi {\rm i} \tau),\quad \tilde{q}: = \exp(-2\pi {\rm i} /(3\tau))
\end{gather}
where $\omega_3=\exp(2\pi {\rm i}/3)$ is a primitive cube root of unity.

Taking a logarithm, we were able to simplify this result, obtaining
\begin{equation}
\log\Big[\mbox{Tr}(\rho_{\mathbb{P}^2})\Big] =
- \frac{1}{2}\log(3^{\frac{5}{2}}\tau) -\frac{\pi {\rm i}}{4} + \frac{3}{2} \left(\widetilde{\mathcal{L}}_1(\chi_{3,2};\tau) - \sqrt{-3} \mathcal{L}_1(\chi_{3,2};-\tfrac{1}{3\tau}) \right)
\end{equation}
with  the odd character $\chi_{3,2}(n)=\pm1$ for $n=\pm1\text{ mod }3$ and $\chi_{3,2}(n)=0$ otherwise.

At large or small coupling  $y=-{\rm i}\tau$ this requires resummation of an infinite number of perturbative terms, 
with an exponentially suppressed tail in 
\begin{equation}\left( \begin{matrix} 
\mathcal{L}_{1}(\chi_{3,2};\tau) \\ 
{\widetilde{\mathcal{L}}}_{1}(\chi_{3,2};\tau ) 
\end{matrix} \right) = \left(\begin{matrix}
\mathcal{S}_\mp\left[ \mathcal{L}^{{\rm Pert}}_{1} \right](\chi_{3,2};\tau ) \vspace{0.1cm}\\ 
\mathcal{S}_\mp\left[ \widetilde{\mathcal{L}}^{{\rm Pert}}_{1} \right](\chi_{3,2};\tau)
\end{matrix}\right) \pm {\rm  i}  \left(\begin{matrix} 0 & \frac{1}{\sqrt{3}} \\ \sqrt{3} & 0\end{matrix}\right)
\left( \begin{matrix} \mathcal{L}_{1}({\chi}_{3,2};-\frac{1}{3\tau}) \\ 
{\widetilde{\mathcal{L}}}_{1}({\chi}_{3,2};-\frac{1}{3\tau} )
\end{matrix} \right)
\end{equation}
resolving the ambiguity of directional Borel resummation, indicated by $S_\mp$.

\subsection{Spectral trace of local $\mathbb{P}^{m,n}$}

Here we need to consider characters with conductor $N=m+n+1$.
With $(m,n)=(2,1)$, we encounter the odd character 
$\chi_{4,3}(n)=\pm1$ for $n=\pm1~\text{mod}~4$ and $\chi_{4,3}(n)=0$ otherwise.
The logarithm of the spectral trace involves the combination
\begin{equation}
\widetilde{\mathcal{L}}_1(\chi_{4,3};\tau)-\sqrt{-4} {\mathcal{L}}_1(\chi_{4,3};-\tfrac{1}{4\tau})
\end{equation}
and hence directional Borel resummation is required at both large and small coupling.

With $(m,n)=(3,1)$, or $(m,n)=(2,2)$, there are three non-principal characters to consider: 
 \begin{center}
\begin{tabular}{ |c|r|r|r|r|} 
\hline
 $n\,\rm{mod}\,5$ & 1 & 2 & 3 & 4 \\
\hline
$\chi_{5,2}(n)$ &1 & ${ \rm i}$ & $-{\rm i}$ & $-1$\\
$\chi_{5,3}(n)$ &1 & $-{\rm i}$ & ${ \rm i}$ & $-1$\\
$\chi_{5,4}(n)$ &1 & $-1$ & $-1$ & 1\\
\hline
\end{tabular}
\end{center}
The odd quartic characters give contributions that require Borel resummation, while
the even quadratic character gives a contribution 
\begin{equation}
\widetilde{\mathcal{L}}_1(\chi_{5,4};\tau) - \sqrt{5}{\mathcal{L}}_1(\chi_{5,4},-\tfrac{1}{5\tau})
\end{equation}
with perturbative expansions that terminate at both large and small coupling.
Yet a non-perturbative tail persists, like the grin of Lewis Carroll's Cheshire cat at sunrise.

\raggedright

\end{document}